# A Near-optimal Algorithm for Edge Connectivity-based Hierarchical Graph Decomposition

Lijun Chang
The University of Sydney
Lijun.Chang@sydney.edu.au

November 25, 2017

## Abstract

Driven by many applications in graph analytics, the problem of computing $k$-edge connected components ($k$-ECCs) of a graph $G$ for a user-given $k$ has been extensively studied recently. In this paper, we investigate the problem of constructing the hierarchy of edge connectivity-based graph decomposition, which compactly represents the $k$-ECCs of a graph for all possible $k$ values. This is based on the fact that each $k$-ECC is entirely contained in a $(k$-$1)$-ECC. In contrast to the existing approaches that conduct the computation either in a bottom-up or a top-down manner, we propose a binary search-based framework which invokes a $k$-ECC computation algorithm as a black box. Let $\mathsf{T}_{\mathsf{KECC}}(G)$ be the time complexity of computing all $k$-ECCs of $G$ for a specific $k$ value. We prove that the time complexity of our framework is $\mathcal{O}\big((\log \delta(G)) \times \mathsf{T}_{\mathsf{KECC}}(G)\big)$, where $\delta(G)$ is the degeneracy of $G$ and equals the maximum value among the minimum vertex degrees of all subgraphs of $G$. As $\delta(G)$ is typically small for real-world graphs, this time complexity is optimal up to a logarithmic factor.

## 1 Introduction

In many real applications, data and their relationships can be modelled as a graph $G = (V, E)$, where vertices in $V$ represent entities of interest and edges in $E$ represent relationships between entities. With the proliferation of graph applications, such as social networks, information networks, web search, collaboration networks, E-commerce networks, communication networks, and biology, research efforts have been devoted towards many fundamental problems in managing and analyzing graph data. Among them, the problem of computing all $k$-*edge connected components* ($k$-ECCs) of a graph $G$ for a user-given $k$ has been extensively studied recently, *e.g.*, in [2, 5, 16, 20]. A $k$-ECC of $G$ is a maximal subgraph $g$ of $G$ such that $g$ is $k$-*edge connected* (i.e., the resulting graph is still connected after the removal of any $k-1$ edges from $g$).

In this paper, we investigate the problem of constructing the hierarchy of Edge Connectivity-based graph decomposition, abbreviated as ECo-decomposition, which compactly represents the $k$-ECCs of a graph for all possible $k$ values. This is based on the following facts of $k$-ECCs [3]: (1) each $k$-ECC is a vertex-induced subgraph; (2) all $k$-ECCs for a specific $k$ value are dis-joint; and (3) each $k$-ECC is entirely contained in a $(k$-$1)$-ECC. Thus, all $k$-ECCs for different $k$ values form a *hierarchy*. Computing the ECo-decomposition of a graph has many applications, *e.g.*, see [17].

**Compute the ECo-decomposition of a Graph.** It is shown in [3] that, after computing the steiner connectivity for all edges of $G$, the hierarchy of the ECo-decomposition of $G$ can be constructed in $\mathcal{O}(m)$ time where $m$ is the number of edges of $G$. The *steiner connectivity* of edge $(u,v)$, denoted $sc(u,v)$, is defined as the largest $k$ such that a $k$-ECC of $G$ contains $(u,v)$. As a result, the main problem of ECo-decomposition is to efficiently compute the steiner connectivity for all edges of $G$.

As the set of edges of $k$-ECCs of $G$ that are not part of $(k{+}1)$-ECCs of $G$ are exactly the set of edges with steiner connectivity $k$, the steiner connectivities of all edges of $G$ can be obtained by computing the $k$-ECCs of $G$ for all possible $k$ values. A bottom-up approach is proposed in [3], which computes $k$-ECCs of $G$ for all possible $k$ values in increasing $k$ value order; we denote it as ECo-BU. Moreover, ECo-BU has an optimization that the input to $k$-ECC computation for a given $k$ is not $G$ but the set of $(k$-$1)$-ECCs of $G$ which has already been obtained in the previous iteration. Similarly, a top-down approach is proposed in [17], which computes $k$-ECCs of $G$ for all possible $k$ values in decreasing $k$ value order, and moreover, when computing $k$-ECCs of $G$ for a given $k$, each $(k{+}1)$-ECC of $G$ which is obtained in the previous iteration is contracted into a single super-vertex; we denote this approach as ECo-TD.

It is easy to see that the time complexities of both ECo-BU and ECo-TD are $\mathcal{O}\big(\delta(G) \times \mathsf{T}_{\mathsf{KECC}}(G)\big)$, where $\mathsf{T}_{\mathsf{KECC}}(G)$ is the time complexity of computing all $k$-ECCs of $G$ for a given $k$, and $\delta(G)$ is the *degeneracy* of $G$ and equals the maximum value among the minimum vertex degrees of all subgraphs of $G$. This is because the largest $k$ that $G$ contains a $k$-ECC is at most $\delta(G)$. That is, there is no $(\delta(G){+}1)$-ECC in $G$, since the minimum vertex degree of a $k$-ECC is at least $k$. Thus, the $k$-ECC computation algorithm is invoked for at most $\delta(G)$ times, and the time complexity follows.

**Our Contributions.** In this paper, we propose a binary search-based framework, denoted ECo-BS, to compute the stener connectivities of all edges of $G$ in $\mathcal{O}\big((\log \delta(G)) \times \mathsf{T}_{\mathsf{KECC}}(G)\big)$ time. The general idea is to conduct a binary search for all possible values of $k$. Moreover, we design computation sharing techniques for $k$-ECC computation of different $k$ values. That is, after computing all $k$-ECCs of $G$ for a chosen $k$, we generate two graphs from $G$: $G_1$ is the set of all $k$-ECCs and is feed as input to $k$-ECC computations for larger $k$ values; $G_2$ is obtained from $G$ by contracting each $k$-ECC into a single super-vertex and is feed as input to $k$-ECC computations for smaller $k$ values. As a result, we are able to prove that ECo-BS runs in $\mathcal{O}\big((\log \delta(G)) \times \mathsf{T}_{\mathsf{KECC}}(G)\big)$ time. Note that, $\delta(G)$ is bounded by $\lceil \sqrt{2m+n} \rceil$ in the worst case, and is small for real-world graphs. Thus, ECo-BS is optimal up to a logarithmic factor.

**Related Work.** We categorize the related works as follows.

*$k$-ECC Computation.* In the literature, there are three approaches for computing all $k$-ECCs of a graph for a given $k$; that is, cut-based approach [13, 16, 20], decomposition-based approach [5], and randomized approach [2]. In this paper, we use the existing $k$-ECC computation algorithm as a black box.

*Dense Subgraph Extraction.* Efficient techniques for computing all maximal *cliques* and *quasi-cliques* of a graph are presented in [4, 7] and [18], respectively. Problems of efficiently computing other dense subgraphs, including $k$-core [6], DN-subgraph [15], triangle $k$-core motifs [19], etc., have also been recently investigated. Nevertheless, due to inherently different problem natures, these techniques are inapplicable to compute ECo-decomposition of a graph.

*Edge Connectivity Computation.* Efficiently computing edge connectivities between vertex-pairs has been studied in graph theory [9], which can be achieved by the maximum flow techniques [8]. The state-of-the-art algorithms compute exact maximum flow in $O(n \times m)$ time [12] and approximate maximum flow in almost linear time to $m$ [11, 14]. To efficiently process vertex-to-vertex edge connectivity queries, index structures have also been developed in [1] and [10]. Note that, the steiner connectivities computed in this paper is different from the edge connectivities computed in the literature. That is, edge connectivity measures the connectivity between two vertices in the input graph $G$, while steiner connectivity measures the connectivity of the most tightly connected subgraph that contains the edge. Thus, these techniques cannot be used to compute ECo-decomposition of a graph.

## 2 Preliminaries

In this paper, we focus on an *undirected graph* $G = (V, E)$, where $V$ is the set of vertices and $E$ is the set of edges. We denote the number of vertices and the number of edges in $G$ by $n$ and $m$, respectively. Given a vertex subset $V_s \subseteq V$, the *vertex-induced subgraph* $G[V_s]$ by $V_s$ is a subgraph $G[V_s] = (V_s, E_s)$ of $G$ with $V_s$ as its vertex set such that $E_s$ consists of only the edges in $G$ with both endpoints in $V_s$; that is, $G[V_s] = (V_s, \{(u,v) \in E \mid u,v \in V_s\})$.

*Definition 2.1:* (**$k$-edge Connected [9]**) A graph $G$ is *$k$-edge connected* if the remaining graph is still connected after the removal of any $(k-1)$ edges from $G$.

*Definition 2.2:* (**$k$-edge Connected Component [2, 5]**) Given a graph $G$, a subgraph $g$ of $G$ is a *$k$-edge connected component* ($k$-ECC) of $G$ if 1) $g$ is $k$-edge connected, and 2) any supergraph in $G$ of $g$ is not $k$-edge connected.

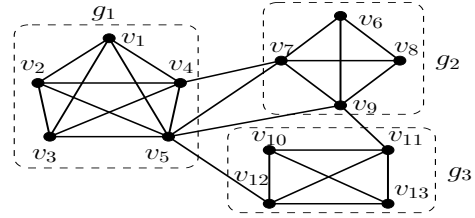

Figure 1: An example graph

For example, the graph $G$ in Figure 1 is 2-edge connected, the subgraph $g_1$ is a 4-ECC, and $g_3$ is a 3-ECC. However, $g_2$ is not a 3-ECC, since $g_1 \cup g_2$ is also 3-edge connected; $g_1 \cup g_2$ is a 3-ECC. Here, $g_1 \cup g_2$ denotes the union of $g_1$, $g_2$, which also includes the edges between vertices of $g_1$ and vertices of $g_2$.

**Properties of $k$-ECCs of a graph [3].**
- Each $k$-ECC is a vertex-induced subgraph.
- The set of all $k$-ECCs of $G$ for a given $k$ is disjoint.
- Each $k$-ECC of $G$ is entirely contained in a $(k\text{-}1)$-ECC of $G$.

As a result, the $k$-ECCs of $G$ for all possible $k$ values form a *hierarchy*.

**Problem 1.** Given an undirected graph $G$, in this paper we study the problem of efficiently constructing the hierarchy of edge connectivity-based graph decomposition of $G$, abbreviated as ECo-decomposition of $G$, which compactly represents all $k$-ECCs of $G$ for all possible $k$ values.

### 2.1 Steiner Connectivity

*Definition 2.3:* (**Steiner Connectivity**) Given a graph $G$, the steiner connectivity of an edge $(u,v)$, denoted $sc(u,v)$, is the largest $k$ such that a $k$-ECC of $G$ contains $u$ and $v$.

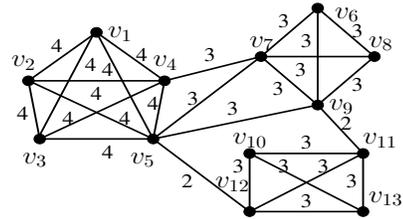

Figure 2: Steiner connectivities

In Figure 2, the steiner connectivity of every edge is shown beside the edge; for example, $sc(v_1, v_4) = 4$.

It is shown in [3] that the hierarchy can be constructed in $\mathcal{O}(m)$ time if the steiner connectivities of all edges of $G$ are given. Thus our main problem becomes efficiently computing the steiner connectivity of all edges of $G$.

**Problem 2.** Given an undirected graph $G$, in this paper we study the problem of efficiently computing the steiner connectivity of all edges of $G$.

# 3 Compute Steiner Connectivities

The existing algorithms compute the steiner connectivities of all edges of $G$ by computing $k$-ECCs of $G$ for all possible $k$ values either in a bottom-up (*i.e.*, from small to large $k$ values) or a top-down (*i.e.*, from large to small $k$ values) manner, based on Lemma 3.1.

**Lemma 3.1:** *The set of edges of $G$ with steiner connectivity $k$ is exactly the set of edges of $k$-ECCs of $G$ that are not part of $(k+1)$-ECCs of $G$.*

Moreover, the time complexities of the existing algorithms can be easily proved based on Lemma 3.2

**Lemma 3.2:** *Let $k_{max}(G)$ be the largest $k$ such that $G$ contains a $k$-ECC. Then, we have $k_{max}(G) \leq \delta(G)$, where $\delta(G)$ is the degeneracy of $G$ and equals the maximum value among the minimum vertex degrees of all subgraphs of $G$.*

**Proof.** We prove the lemma by contradiction. Assume $k_{max}(G) > \delta(G)$, and let $g$ be a $(k_{max}(G))$-ECC of $G$. Then, the minimum vertex degree of $g$ is at least $k_{max}(G)$ which contradicts that the degeneracy of $G$ is $\delta(G)$ which is smaller than $k_{max}(G)$. Thus, the lemma holds. □

## 3.1 A Bottom-Up Approach

Based on Lemma 3.1, a bottom-up approach is proposed in [3] which computes the steiner connectivities of all edges in $\mathcal{O}(\delta(G) \times \mathsf{T}_{\mathsf{KECC}}(G))$ time, where $\mathsf{T}_{\mathsf{KECC}}$ is the time complexity of a $k$-ECC computation algorithm for a given $k$. The pseudocode is shown in Algorithm 1, denoted ECo-BU.

---

**Algorithm 1:** ECo-BU

**Input**: A graph $G$
**Output**: $sc(u, v)$ for each edge $(u, v)$ in $G$

1 $\phi_1(G) \leftarrow \{G\}; k \leftarrow 1;$
2 **while** $\phi_k(G) \neq \emptyset$ **do**
3      $k \leftarrow k + 1; \phi_k(G) \leftarrow \emptyset;$
4      **for each** *graph $g$ of size at least $2$ in $\phi_{k-1}(G)$* **do**
5          $\phi_k(G) \leftarrow \phi_k(G) \cup \mathsf{ComputeKECCs}(g, k);$
6          Assign $sc(u, v)$ to be $k - 1$ for each edge $(u, v)$ removed during the computation at Line 5;

---

**Example 3.1:** Consider the graph in Figure 1. $\phi_2(G) = \{G\}$. $\phi_3(G) = \{g_1 \cup g_2, g_3\}$; $(v_5, v_{12})$ and $(v_9, v_{11})$ are removed, thus $sc(v_5, v_{12}) = 2$ and $sc(v_9, v_{11}) = 2$. In computing $\phi_4(G)$, all edges in $G$ except those in $g_1$ are removed; therefore, these newly removed edges $(u, v)$ have $sc(u, v) = 3$. Finally, the edges $(u', v')$ in $g_1$ are removed and have $sc(u', v') = 4$. The steiner connectivities of all edges are shown in Figure 2. □

## 3.2 A Top-Down Approach

Recently, a top-down approach is proposed in [17] but in the context of I/O-efficient algorithm. We modify it to be a main memory algorithm and shown in Algorithm 2, denoted ECo-TD. It is easy to see that the time complexity of ECo-TD is $\mathcal{O}(\delta(G) \times \mathsf{T}_{\mathsf{KECC}}(G))$.

---

**Algorithm 2:** ECo-TD

**Input**: A graph $G$
**Output**: $sc(u, v)$ for each edge $(u, v)$ in $G$

1 Let $k$ be the degeneracy $\delta(G)$ of $G$;
2 **while** $k > 0$ **do**
3      $\phi_k(G) \leftarrow \mathsf{ComputeKECCs}(G, k);$
4      Assign $sc(u, v)$ to be $k$ for each edge $(u, v)$ in the subgraphs of $\phi_k(G);$
5      Contract each subgraph that is in $\phi_k(G)$ into a single super-vertex in $G;$

---

Note that, the top-down algorithm proposed in [17] uses a looser upper bound than $\delta(G)$, since they focus on I/O-efficient algorithms. Thus, directly modifying the top-down algorithm in [17] to run in main memory has a higher time complexity than $\mathcal{O}(\delta(G) \times \mathsf{T}_{\mathsf{KECC}}(G))$.

# 4 A Binary Search-based Framework

In this section, we propose a new binary search-based framework to compute the steiner connectivities of all edges of a graph in a near-optimal time. Let $\overline{k}_{max}$ be an upper bound of $k_{max}$. Note that, although in this paper we use $\delta(G)$ as $\overline{k}_{max}$, we present our framework in a more general form. The general idea is that, instead of computing $k$-ECCs for $k$ varying sequentially from $\overline{k}_{max}$ to $2$ or the other way around, we conduct a binary search on the interval $[2, \overline{k}_{max}]$ of $k$ values. Specifically, we first compute the $\lfloor \frac{2+\overline{k}_{max}}{2} \rfloor$-ECCs of the input graph $G$, based on which we obtain two graphs $G_1$ and $G_2$. Here, $G_1$ is the computed set of $\lfloor \frac{2+\overline{k}_{max}}{2} \rfloor$-ECCs, and $G_2$ is obtained from $G$ by contracting every $\lfloor \frac{2+\overline{k}_{max}}{2} \rfloor$-ECC into a super-vertex. Then, we recursively solve the problem for $k$ values in the interval of $[\lfloor \frac{2+\overline{k}_{max}}{2} \rfloor + 1, \overline{k}_{max}]$ on $G_1$, and also the problem for $k$ values in the interval of $[2, \lfloor \frac{2+\overline{k}_{max}}{2} \rfloor - 1]$ on $G_2$. The pseudocode is shown in Algorithm 3, which is self explanatory.

**Theorem 4.1:** *Invoking Algorithm 3 with the input interval $[2, \overline{k}_{max}]$ correctly computes the steiner connectivities of all edges of a graph.*

**Proof.** Firstly, we prove that for each invocation of Algorithm 3 with inputs $G'$ and $[L, H]$, all connected components of $G'$ are $(L-1)$-edge connected and there is no $(H+1)$-ECC of $G'$. Initially, this property trivially holds for the first invocation of Algorithm 3. Now, we show that, if this property holds for the inputs $G$ and $[L, H]$, then it also holds for the subroutines invoked at Line 6 and Line 11. For Line 6, each connected component of $G_1$ is a $M$-ECC of $G$ and is thus $M$-edge connected, and moreover, $G_1$ has no $(H+1)$-ECC since it is a subgraph

**Algorithm 3:** ECo-BS

**Input**: A graph $G$, and an interval $[L, H]$ of steiner connectivity values with $L \leq H$
**Output**: $sc(u,v)$ for each edge $(u,v)$ in $G$

1 $M \leftarrow \lfloor \frac{L+H}{2} \rfloor$;
2 $G_1 \leftarrow$ ComputeKECCs($G, M$);  /* Compute $k$-ECCs of $G$ for a given $k = M$ */;
3 **if** $M = H$ **then**
4      Assign $sc(u,v)$ to be $H$ for each edge $(u,v)$ in $G_1$;
5 **else**
6      ECo-BS($G_1, [M+1, H]$);
7 **if** $M = L$ **then**
8      Assign $sc(u,v)$ to be $(L-1)$ for each edge $(u,v)$ removed during the computation at Line 2;
9 **else**
10      $G_2 \leftarrow$ the graph obtained from $G$ by contracting each subgraph that is a connected component of $G_1$ into a super-vertex;
11      ECo-BS($G_2, [L, M-1]$);

of $G$; thus, the property holds when going into the subroutine at Line 6. For Line 11, every connected component of $G_2$ is $(L-1)$-edge connected since it is obtained from a connected component of $G$ by contracting each $M$-ECC into a super-vertex where $M > L$. Also, there is no $M$-ECC in $G_2$ since every $M$-ECC is contracted into a super-vertex. Hence, the property holds when going into the subroutine at Line 11.

Now, we prove the theorem by inducting on the length of the interval $[L, H]$; that is, $len = H - L + 1$. For the base case that $len = 1$ (i.e., $L = H$), we have $M = L = H$. We assign $sc(u,v)$ to be $M$ for each edge $(u,v)$ in the $M$-ECCs of $G$, and assign $sc(u,v)$ to be $M - 1$ for each edge removed during computing the $M$-ECCs of $G$. Hence, the algorithm is correct, since every connected component of $G$ is $(L-1)$-edge connected and there is no $(H+1)$-ECC in $G$. Now, assume that the algorithm is correct for $H - L + 1 \leq r$ with $r \geq 1$, we prove that the algorithm is also correct for $H - L + 1 = r + 1$. Let $M = \lfloor \frac{L+H}{2} \rfloor$, we have $L \leq M < H$. After computing the $M$-ECCs of $G$, we partition the set of edges of $G$ into two sets: the set of edges in the $M$-ECCs of $G$ and the set of edges not in the $M$-ECCs of $G$, which correspond to the set of edges in $G_1$ and the set of edges in $G_2$, respectively. It is easy to verify that for each edge $(u,v)$ in $G$, $(u,v)$ is in $G_1$ and $sc(u,v)$ equals the steiner-connectivity of $(u,v)$ in $G_1$ if $sc(u,v) \geq M$, and $(u,v)$ is in $G_2$ and $sc(u,v)$ equals the steiner-connectivity of $(u,v)$ in $G_2$ if $sc(u,v) < M$; these two cases are correctly computed at Line 6 and Line 11, respectively. Thus, the theorem holds. $\square$

**Theorem 4.2:** *The time complexity of Algorithm 3 with the input interval $[2, \overline{k}_{max}]$ is $\mathcal{O}\big((\log \overline{k}_{max}) \times \mathsf{T}_{\mathsf{KECC}}(G)\big)$.*

**Proof.** We prove the theorem by inducting on the length of the interval $[L, H]$ in the input of Algorithm 3. Obviously, when $L = H$, the time complexity of Algorithm 3 is $\mathcal{O}(\mathsf{T}_{\mathsf{KECC}}(G)) = \mathcal{O}((\log(H-L+2)) \times \mathsf{T}_{\mathsf{KECC}}(G))$; thus, the theorem holds when $L = H$. Now, we assume the theorem holds for all intervals of length $len$ (i.e., $H - L + 1 = len$), we prove that it also holds for intervals of length $len + 1$. Given an arbitrary interval $[L, H]$ with $H - L = len$, Line 2 takes time $\mathcal{O}(\mathsf{T}_{\mathsf{KECC}}(G))$, and from the induction we have that the recursion of Lines 3–6 takes time $\mathcal{O}\big((\log(H - M + 1)) \times \mathsf{T}_{\mathsf{KECC}}(G_2)\big)$ and the recursion of Lines 7–11 takes time $\mathcal{O}\big((\log(M - L + 1)) \times \mathsf{T}_{\mathsf{KECC}}(G_1)\big)$. Without loss of generality, we assume that $H - M = M - L$. Then, the total time complexity of Algorithm 3 is $\mathcal{O}\big(\mathsf{T}_{\mathsf{KECC}}(G) + (\log(H - M + 1)) \times (\mathsf{T}_{\mathsf{KECC}}(G_1) + \mathsf{T}_{\mathsf{KECC}}(G_2))\big) = \mathcal{O}\big((\log(H - M + 1) + 1) \times \mathsf{T}_{\mathsf{KECC}}(G)\big) = \mathcal{O}((\log(H - L + 2)) \times \mathsf{T}_{\mathsf{KECC}}(G))$, where the first equality holds since $E(G_1) \cup E(G_2) = E$ and $E(G_1) \cap E(G_2) = \emptyset$, and we assume that $\mathsf{T}_{\mathsf{KECC}}(G)$ is linear or super-linear to $E(G)$; here $E(G)$ denotes the set of edges of $G$. Thus, the time complexity of Algorithm 3 with input interval $[2, \overline{k}_{max}]$ is $\mathcal{O}\big((\log \overline{k}_{max}) \times \mathsf{T}_{\mathsf{KECC}}(G)\big)$, and the theorem holds. $\square$

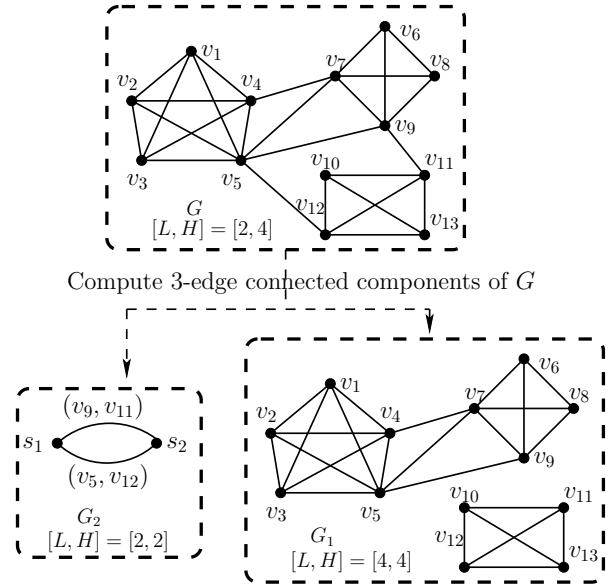

Figure 3: Running example of Algorithm 3

**Example 4.1:** Consider the graph $G$ in Figure 1 which is also shown in the top part of Figure 3. Assume $\overline{k}_{max}$ is computed as 4. Then, we compute the steiner connectivities for all edges of $G$ by invoking Algorithm 3 with input $G$ and $[L, H] = [2, 4]$. That is, we compute the $\lfloor \frac{2+4}{2} \rfloor$-ECCs of $G$, and obtain the subgraph induced by $S_1 = \{v_1, v_2, \ldots, v_9\}$ and the subgraph induced by $S_2 = \{v_{10}, \ldots, v_{13}\}$. As $L < k = 3 < H$, we will continue the computation for the graph $G_1$ and $G_2$ with intervals $[2, 2]$ and $[4, 4]$, respectively.

The graph $G_1$ consists of the two 3-ECCs of $G$, and is shown in the lower right part of Figure 3. We compute the 4-ECCs of $G_1$, and obtain the subgraph induced by vertices $\{v_1, v_2, \ldots, v_5\}$. Thus, all edges among vertices $\{v_1, v_2, \ldots, v_5\}$ have steiner connectivities 4, and other edges have steiner connectivities 3. The graph $G_2$ is obtained by contracting each of $S_1$ and $S_2$ into a super-vertex, and is shown in the lower left part of Figure 3; in $G_2$, there are two parallel edges between $s_1$ and $s_2$, corresponding to edges $(v_9, v_{11})$ and ($v_5, v_{12}$), respectively. As $G_2$ is 2-edge connected, the steiner-connectivities of ($v_9, v_{11}$) and ($v_5, v_{12}$) are 2. The result is the same as that computed by Algorithm 1. □

Thus, by let $\overline{k}_{max} = \delta(G)$ which can be computed in $\mathcal{O}(m)$ time, the time complexity of ECo-BS is $\mathcal{O}\big((\log \delta(G)) \times \mathsf{T}_{\mathsf{KECC}}(G)\big)$.

# References

[1] C. C. Aggarwal, Y. Xie, and P. S. Yu. Gconnect: A connectivity index for massive disk-resident graphs. *PVLDB*, 2(1):862–873, 2009.

[2] T. Akiba, Y. Iwata, and Y. Yoshida. Linear-time enumeration of maximal k-edge-connected subgraphs in large networks by random contraction. In *Proc. CIKM'13*, 2013.

[3] L. Chang, X. Lin, L. Qin, J. X. Yu, and W. Zhang. Index-based optimal algorithms for computing steiner components with maximum connectivity. In *Proc. of SIGMOD'15*, 2015.

[4] L. Chang, J. X. Yu, and L. Qin. Fast maximal cliques enumeration in sparse graphs. *Algorithmica*, 66(1), 2013.

[5] L. Chang, J. X. Yu, L. Qin, X. Lin, C. Liu, and W. Liang. Efficiently computing k-edge connected components via graph decomposition. In *Proc. SIGMOD'13*, 2013.

[6] J. Cheng, Y. Ke, S. Chu, and M. T. Özsu. Efficient core decomposition in massive networks. In *Proc. of ICDE'11*, 2011.

[7] J. Cheng, Y. Ke, A. W.-C. Fu, J. X. Yu, and L. Zhu. Finding maximal cliques in massive networks by h*-graph. In *Proc. of SIGMOD'10*, 2010.

[8] T. H. Cormen, C. Stein, R. L. Rivest, and C. E. Leiserson. *Introduction to Algorithms*. McGraw-Hill Higher Education, 2001.

[9] A. Gibbons. *Algorithmic Graph Theory*. Cambridge University Press, 1985.

[10] R. E. Gomory and T. C. Hu. Multi-Terminal Network Flows. *Journal of the Society for Industrial and Applied Mathematics*, 9(4), 1961.

[11] J. A. Kelner, Y. T. Lee, L. Orecchia, and A. Sidford. An almost-linear-time algorithm for approximate max flow in undirected graphs, and its multicommodity generalizations. In *Proc. of SODA'13*, 2013.

[12] J. B. Orlin. Max flows in o(nm) time, or better. In *Proc. of STOC'13*, 2013.

[13] A. N. Papadopoulos, A. Lyritsis, and Y. Manolopoulos. Skygraph: an algorithm for important subgraph discovery in relational graphs. *Data Min. Knowl. Discov.*, 17(1), Aug. 2008.

[14] J. Sherman. Nearly maximum flows in nearly linear time. In *Proc. of FOCS'13*, 2013.

[15] N. Wang, J. Zhang, K.-L. Tan, and A. K. H. Tung. On triangulation-based dense neighborhood graph discovery. *Proc. VLDB Endow.*, 4(2), Nov. 2010.

[16] X. Yan, X. J. Zhou, and J. Han. Mining closed relational graphs with connectivity constraints. In *Proc. of KDD'05*, 2005.

[17] L. Yuan, L. Qin, X. Lin, L. Chang, and W. Zhang. I/O efficient ECC graph decomposition via graph reduction. *VLDB J.*, 26(2), 2017.

[18] Z. Zeng, J. Wang, L. Zhou, and G. Karypis. Out-of-core coherent closed quasi-clique mining from large dense graph databases. *ACM Trans. Database Syst.*, 32(2), June 2007.

[19] Y. Zhang and S. Parthasarathy. Extracting analyzing and visualizing triangle k-core motifs within networks. In *Proc. of ICDE'12*, 2012.

[20] R. Zhou, C. Liu, J. X. Yu, W. Liang, B. Chen, and J. Li. Finding maximal k-edge-connected subgraphs from a large graph. In *Proc. of EDBT'12*, 2012.